\documentclass[article]{jss}
\usepackage{thumbpdf,lmodern}
\usepackage{framed}
\usepackage{xcolor}
\usepackage{listings}
\usepackage{inconsolata}
\usepackage{amsmath}
\usepackage{subfig}
\usepackage{lineno}

\definecolor{bluekeywords}{rgb}{0,0,1}
\definecolor{greencomments}{rgb}{0,0.5,0}
\definecolor{redstrings}{rgb}{0.64,0.08,0.08}
\definecolor{xmlcomments}{rgb}{0.5,0.5,0.5}
\definecolor{types}{rgb}{0.17,0.57,0.68}

\newcommand{\matrixcl}{\code{matrix\_cl}{}}
\newcommand{\kernelcl}{\code{kernel\_cl}{}}
\newcommand{\outbuffer}{\code{out\_buffer}{}}
\newcommand{\inbuffer}{\code{in\_buffer}{}}

\author{Rok \v{C}e\v{s}novar \\ University of Ljubljana
\And Steve Bronder\\University of Columbia
   \AND Davor Sluga \\University of Ljubljana
   \And Jure Dem\v{s}ar\\University of Ljubljana
   \And Tadej Ciglari\v{c}\\University of Ljubljana
   \AND Sean Talts \\Columbia University
   \And Erik \v{S}trumbelj\\University of Ljubljana
   }
\Plainauthor{Rok \v{C}e\v{s}novar, Steve Bronder, Davor Sluga, Jure Dem\v{s}ar, Tadej Ciglari\v{c}, Sean Talts, Erik \v{S}trumbelj}

\title{GPU-based Parallel Computation Support for \proglang{Stan}}
\Plaintitle{GPU-based Parallel Computation Support for Stan}
\Shorttitle{GPU-based Parallel Computation Support for \proglang{Stan}}

\Abstract{This paper details an extensible \proglang{OpenCL} framework that allows \proglang{Stan} to utilize heterogeneous compute devices. It includes GPU-optimized routines for the Cholesky decomposition, its derivative, other matrix algebra primitives and some commonly used likelihoods, with more additions planned for the near future. \proglang{Stan} users can now benefit from large speedups offered by GPUs with little effort and without changes to their existing \proglang{Stan} code. We demonstrate the practical utility of our work with two examples -- logistic regression and Gaussian Process regression.}

\Keywords{statistical inference, Bayesian statistics, automatic differentiation, parallel computation, GPU, \proglang{Stan}}
\Plainkeywords{Statistical inference, Bayesian statistics, automatic differentiation, parallel computation, GPU, Stan}

\Address{
  Rok \v{C}e\v{s}novar, Davor Sluga, Jure Dem\v{s}ar, Tadej Ciglari\v{c}, Erik \v{S}trumbelj\\
  University of Ljubljana\\
  Faculty of Computer and Information Science\\
  E-mail: \email{erik.strumbelj@fri.uni-lj.si}
  
  Steve Bronder\\
  ISERP\\
  Columbia University\\
  E-mail: \email{sab2287@columbia.edu}
  
  Sean Talts\\
  ISERP\\
  Columbia University\\
  E-mail: \email{sean.talts@gmail.com}
  
}

\begin{document}

\section{Introduction}\label{sec:intro}

\proglang{Stan} is an open-source probabilistic programming language for Bayesian modeling and inference \citep{Carpenter2017}. It has become the system of choice for statisticians and scientists as well as the reference project for Bayesian inference. Hundreds of research papers using \proglang{Stan} are published every year, ranging from cognitive anthropology and the structure of gravitational clusters to clinical trial design, and sports. Dozens of tools utilize \proglang{Stan} such as \pkg{rstanarm} \citep{gabry2016rStanarm}, \pkg{brms} \citep{burkner2017brms}, and Facebook's forecasting tool \pkg{Prophet} \citep{taylor2018forecasting}.

There exist many other languages and software tools similar to \proglang{Stan}. Some focus more on statistical inference, while others focus more on machine learning and deep learning. To name just a few of the most popular:  \pkg{Edward}/\pkg{Edward2} (\pkg{TensorFlow}) \citep{Tran2017} and \pkg{PyMC3} \citep{Salvatier2016} (\pkg{Theano}) \citep{bergstra2010theano}, \pkg{Pyro} \citep{bingham2019pyro} (\pkg{PyTorch}) \citep{paszke2017automatic}, and \pkg{MxNet} \citep{chen2015mxnet}.

\proglang{Stan} has three distinct components: a probabilistic programming language, the \pkg{Stan Math} library that supports automatic differentiation, and algorithms for inference. The main advantages of \proglang{Stan} are a rich math library and state-of-the-art inference with a variant of Hamiltonian Monte Carlo -- the NUTS (No-U-turn) sampler \citep{Hoffman2014} with some additional modifications \citep{Betancourt2017b, Stan2020} -- which makes \proglang{Stan}  suitable for robust fully-Bayesian inference. Moreover, the \proglang{Stan} probabilistic programming language is easier to understand than systems embedded in other languages \citep{Baudart2018}.

\proglang{Stan} has only recently started building out low-level parallelism. While \proglang{Stan} supports threading and MPI to execute disjoint sets in the automatic differentiation expression tree, it did not have support for specialized hardware such as GPUs. An ideal case for GPU based optimization are models based on Gaussian Processes (GP). The computation in GP-based models is, even for moderate input sizes, dominated by computing the inverse of the covariance matrix. The $O(n^3)$ time of this operation also dominates the quadratic costs associated with transferring matrices to and from a GPU. In turn, computing the Cholesky decomposition of the positive definite matrix dominates the computation time of the matrix inverse. Because these costs can be broken up and executed in parallel they make the Cholesky decomposition an ideal target for GPU-based computation.

This paper describes a framework for GPU support in \proglang{Stan} and GPU implementations of the Cholesky decomposition, its derivative, other matrix algebra primitives, and GLM likelihoods with derivatives in the \pkg{Stan Math} library. Unlike most similar libraries, our framework relies on \proglang{OpenCL} 1.2 \citep{StoneOpenCL2010}, so it supports a variety of devices. This includes GPUs of different vendors, multi-core CPUs, and other accelerators.

The integration with \proglang{Stan} is seamless and user friendly - setting a flag moves the computation of supported routines to the GPU, with no need to change \proglang{Stan} code. The API provides experts with a simple way of implementing their GPU kernels, using existing GPU kernels as building blocks. We demonstrate the practical utility of our work -- ease of use and speedups -- with two examples, logistic regression and Gaussian Process regression.

\section[Integrating OpenCL with Stan]{Integrating \proglang{OpenCL} with \proglang{Stan}}\label{sec:opencl}

\proglang{Stan}'s reverse mode automatic differentiation uses the \code{Matrix} type from \pkg{Eigen} \citep{eigenweb} to store data as a matrix of type \code{double} or \proglang{Stan}'s \code{var} type, where the \code{var} holds the value and adjoint used in automatic differentiation. \proglang{Stan} builds up an expression tree used in automatic differentiation and stores all the data needed in the expression tree via its local allocator. When a node's children in the expression graph are disjoint, \proglang{Stan} can utilize \proglang{C++11} threading or MPI to compute the log probability evaluation in parallel. When an operation within a node is expensive, \proglang{Stan} can use the \proglang{OpenCL} backend to parallelize the operation on the node. 

\subsection[OpenCL base]{\proglang{OpenCL} base}\label{sec:basecl}

\proglang{Stan}'s \proglang{OpenCL} backend uses a single context to receive data and routines from individual nodes in the expression tree. Ensuring there is only one context and queue per device for the life of the program makes context management simpler. The implementation of the \proglang{OpenCL} context which manages the device queue follows the singleton pattern and sits in the class \code{opencl\_context\_base}.

Instead of calling \code{opencl\_context\_base::getInstance().method()}, developers can access the context through a friend adapter class \code{opencl\_context} which provides an API for accessing the base context. If the \proglang{OpenCL} implementation supports asynchronous operations, then the context asynchronously executes kernels. Asynchronous operations are particularly useful in conjunction with threading as the individual threads will be able to enqueue operations, which will execute while threads do other calculations using CPU.

\subsection[Matrix class]{Matrix class}\label{sec:matclass}

The base matrix class \matrixcl{} holds the device memory buffer, meta-information on the matrix, and methods for reading and writing event stacks for asynchronous computation. When a kernel receives a \matrixcl{}, the kernel's event is attached to the appropriate read or write event stack. Reading and writing to \proglang{OpenCL} buffers uses the generic \code{enqueue(Write)/(Read)Buffer} methods. Because \pkg{Stan Math} heavily relies on \pkg{Eigen} matrices, constructors and methods are available for easily passing data back and forth.

Developers can pass in \pkg{Eigen} matrices directly to the \matrixcl{} constructor or use the \code{to\_matrix\_cl()} or \code{from\_matrix\_cl()} methods.

\begin{lstlisting}
Eigen::MatrixXd m(2, 2);
m << 1, 2, 3, 4;

matrix_cl A(m);
matrix_cl B(2, 2);

B = to_matrix_cl(m);
Eigen::MatrixXd C = from_matrix_cl(B);
\end{lstlisting}

Similar constructors for the \matrixcl{} class are available for standard vectors \code{std::vector<T>} and arrays of doubles.

\begin{figure}[htp]
\includegraphics[width=8cm]{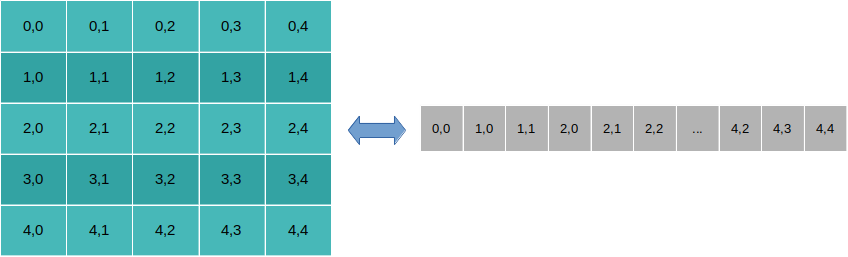}
\centering
\caption{Packing and unpacking a triangular matrix.}
\label{fig:packUnpack}
\end{figure}

We can reduce the data transfers of triangular matrices by only transferring the non-zero parts of the matrix in a packed form. The kernel \texttt{unpack} deals with unpacking the packed form shown on the right-hand side on Figure \ref{fig:packUnpack} to a flat matrix shown on the left-hand side. For lower (upper) triangular matrices, the irrelevant upper (lower) triangular is filled with zeros. The kernel \texttt{pack} packs the flat matrix to packed form for the transfer back to the host's global memory.

\begin{lstlisting}
  matrix_cl L = packed_copy<stan::math::matrix_cl_view::Lower>(L_val_cpu, M_);
\end{lstlisting}

When operating on GPUs, transferring data from host to device and making copies can be the most expensive operations. To reduce the burden of data transfers, the Stan compiler identifies immutable data objects. These objects are copied to the GPU at the start and remain on the GPU until the end of inference.

Note that \code{matrix_cl} supports direct allocation of a matrix of \emph{var} objects (used for parameters for the purpose of automatic differentiation) - this stores two separate matrices, one for values and the other for adjoints, which simplifies the API.

\subsection[Kernel construction]{Kernel construction}\label{sec:kernstruct}

The \proglang{OpenCL} specification demands that strings are used to represent \proglang{OpenCL} kernels. However, having a large number of files comprised of strings is unwieldy and difficult to maintain. \proglang{Stan} wraps its kernels inside of a \code{STRINGIFY} macro, which gives developers access to the standard set of developer tools such as code highlighting, linting, Doxygen \citep{van2008doxygen}, and auto-formatting. This style makes the kernel code easier to maintain compared to having files full of strings. An example of how a developer brings a new kernel into \proglang{Stan}:

\begin{lstlisting}
// Items in between \ cond and \ endcond are ignored by doxygen .
// \ cond
const char * example_kernel_code = STRINGIFY (
// \ endcond
/**
* Example of adding new kernel in Stan
*
* @param [ out] A An example output matrix .
* @param [in] B An example input matrix .
* @param val Some other input value
*/
__kernel void example ( double *A, double *B, int * val ) {
// kernel code ...
}
// \ cond
);
// \ endcond
/**
* See the docs for \ link kernels / example .hpp example () \ endlink
*/
const kernel_cl < out_buffer , in_buffer , int > example (
" example ", example_kernel_code , {" THREAD_BLOCK_SIZE ", 32});
\end{lstlisting}

In the above, a developer uses \code{STRINGIFY} to create a \code{const char*} that holds the kernel code. That string passes into the \kernelcl{} struct templated by the kernel argument types and with arguments giving the name of the kernel, the kernel code, and optional kernel macros they would like to have defined in the kernel. 

Internally, we keep track of \proglang{OpenCL} events via queues on each \matrixcl{} object that we use to conservatively prevent race conditions and provide ordering where necessary. \outbuffer{} and \inbuffer{} are empty structs that we pass as template arguments to configure the kernel during construction to indicate the directionality of each input buffer. At runtime, the kernel will check the correct event queues on its arguments for events it needs to wait for and then attach the event representing the kernel's completion to each buffer's queues correctly. That way we ensure that an operation that, for example, writes to a buffer, is completed before we allow the \proglang{OpenCL} runtime to read from that buffer.

The parameter pack of types in the template for \kernelcl{} are unpacked and passed down as the argument types for the \code{operator()} and down to the template arguments for \proglang{OpenCL}'s \code{make\_kernel} functor. Below is a simplified version of the code used to construct and call the kernel.

\begin{lstlisting}
template <typename... Args>
struct kernel_cl {

  const kernel_functor<to_const_buffer_t<Args>&...> make_functor;

  kernel_cl(const char* name, const std::vector<const char*>& sources,
            const std::map<const char*, int>& options = {})
      : make_functor(name, sources, options) {}
 
  auto operator()(cl::NDRange global_thread_size, cl::NDRange thread_block_size,
                  to_const_matrix_cl_t<Args>&... args) const {
    auto f = make_functor();
    
    const std::vector<cl::Event> kernel_events
        = vec_concat(select_events<Args>(args)...);
    cl::EnqueueArgs eargs(opencl_context.queue(), kernel_events,
                          global_thread_size, thread_block_size);
 
    cl::Event kern_event = f(eargs, get_kernel_args(args)...);

    assign_events<Args...>(kern_event, args...);
    return kern_event;
  }
};
\end{lstlisting}

Note that the meta-programming traits \code{to\_const\_buffer\_t<>} and \code{to\_const\_matrix\_cl\_t<>} override the \inbuffer{} and \outbuffer{} template types in order to propagate a \code{cl::Buffer} or \code{matrix_cl} to subsequent templates and signatures.

In the above code, the \kernelcl{}'s constructor passes the name, kernel code, and kernel options to the kernel functor in the initialization list which compiles the kernel. Kernel arguments declared with \inbuffer{} or \outbuffer{} should be of type \matrixcl{}. When a kernel is called, the events that are in each \matrixcl{}'s read or write stacks are collected depending on whether it was designated as an in or out buffer. The kernel will then wait to execute until the previous events complete. The kernel's event is assigned to each \matrixcl{}'s read and write event stack via \code{assign\_events()} depending on whether it was defined as an \inbuffer{} or an \outbuffer{}.

When the \kernelcl{} struct is constructed, it compiles the kernel and developers call the kernel with

\begin{lstlisting}
matrix_cl foo = //...
matrix_cl goo;
example(cl::NDRange(...), goo, foo, 10);
\end{lstlisting}

Depending on the \code{in/out\_buffer} passed when constructing the kernel, events will be added to the appropriate \matrixcl{} read and/or write event stack. For instance, in the above, \code{goo} in the output and will have the kernel's event attached to it's \code{write\_stack}. While \code{foo} will have the kernel's event attached to its \code{read\_stack}. Later kernel calls that write to \code{foo} will know to wait for all the events in \code{foo}'s \code{read\_stack} and \code{write\_stack} while kernels that use \code{goo} as input will know to wait for the event's in \code{goo}'s \code{write_stack}.

The kernel functions for addition, subtraction, and multiplication are wrapped in their standard operators so users of the \pkg{Stan Math} library can call the operations such as:

\begin{lstlisting}
matrix_cl A = B * C + D - E;
\end{lstlisting}

\section[GPU-optimized routines in the Stan Math library]{GPU-optimized routines in the \pkg{Stan Math} library}\label{sec:routines}

In this section we describe the three GPU-optimized matrix algebra primitives that are currently implemented in \pkg{Stan Math}. They are currently not accessible directly in the \proglang{Stan} language but are used inside existing primitive and reverse functions. Several supporting routines that were used in the implementation of these primitives are described in Section \ref{sec:supportRoutines}.

\subsection[Matrix multiplication]{Matrix multiplication}\label{sec:matrixMul}

Efficient implementations of the general matrix multiplication in \proglang{OpenCL} can be found in various best practices guides \citep{OpenCLBestPractices,CUDABestPractices} and research papers \citep{Matsumoto2014,clBLAST,Nugteren2018}. In \pkg{Stan Math} we implemented general matrix multiplication with optimizations for triangular input matrices, vector and row vector inputs, and multiplication in the form $C=AA^T$. Multiplications of a matrix or its diagonal with a scalar are also supported and explained in Section \ref{sec:supportRoutines}.

We implemented a kernel for general matrix multiplication (\code{GEMM}) that is based on the aforementioned best practices guides and is exposed through the \code{operator*(matrix\_cl\& A, matrix\_cl\& B)} function. Matrix multiplication $C=AB$, where $A$ is $n\times k$ and $B$ is $k\times m$ is executed with a 2D grid of $\frac{n\times m}{\text{WPT}}$ threads, where WPT (Work Per Thread) is an implementation parameter. Thread $(i,j)$ computes the values $C_{i,j}, ..., C_{i,j+\text{WPT}-1}$ of the resulting matrix, therefore performing up to WPT dot products of rows in $A$ and columns in $B$. The default value is WPT = 8 as it gives good overall performance on GPUs of different architectures but it can be tuned for the target GPU. The dot products are done in smaller chunks (tiles) in order to facilitate the use of the GPUs small and fast shared memory. The use of shared memory with tiling is another common approach to optimization in \proglang{OpenCL} and \proglang{CUDA} \code{GEMM} implementations.
When $A$ or $B$ are triangular, threads automatically adapt the start and end of the dot products to avoid unnecessary reads for elements known to be zero. Triangularity information is passed as part of the \code{matrix\_cl} object. Matrix methods utilize this information and set it appropriately for their outputs, if applicable. For example:

\begin{lstlisting}
// triangularity information can be set at object initialization
matrix_cl A(A_cpu);
matrix_cl A_low(A_cpu,matrix_cl_view::Lower);
matrix_cl A_upp(A_cpu,matrix_cl_view::Upper);
matrix_cl B(B_cpu);
matrix_cl B_low(B_cpu,matrix_cl_view::Lower);
matrix_cl b_upp(B_cpu,matrix_cl_view::Upper);

// different underlying multiplication algorithms are executed based on input
matrix_cl C = multiply(A , B);     
matrix_cl C = multiply(A_low, B); 
matrix_cl C = multiply(A, B_upp);
matrix_cl C = multiply(A_upp, B_low);
\end{lstlisting}

\subsubsection{Special case: large $k$}

When $k$ is multiple orders of magnitude larger than $n$ and $m$ and $n\times m$ is small, a small number of threads will compute long dot products thus reducing the occupancy of the GPU. In such cases we instead create $n\times m\times s$ threads, splitting the dot products into $s$ parts, each thread calculating a part of the scalar product. In order to avoid race conditions, we create $s$ copies of the resulting matrix, with the thread's ID in the 3rd dimension determining the copy to write the result to. A separate support kernel then adds these copies together using $n\times m$ threads with thread $(i,j)$ assigned to add the $s$ copies of $C_{i,j}$. Since $n\times m$ is small, these extra copies do not have a large memory footprint. The value $s$ is determined based on the size of the input matrices. This optimization offers up to a 45\% reduction in execution time compared to the \code{GEMM} kernel.

\subsubsection{Special case: $C=AA^T$}

Because $C$ is symmetric the threads where $j<i$ compute the values $C_{i,j}, ..., C_{i,j+\text{WPT}-1}$ and map these values over the diagonal to $C_{j,i}, ..., C_{j+\text{WPT}-1, i}$. Threads where $j>i$ return immediately. This kernel is accessible through \code{multiply\_transpose(matrix\_cl\& A)}.

\subsubsection{Special case: vector inputs}

We also implemented kernels that optimize multiplication of matrices with vectors, row vectors and scalars. The latter is only used as an internal function to implement other primitive function or gradients and is described in Section \ref{sec:supportRoutines}.

The general matrix multiplication kernel is not suitable for cases where the inputs are vector or row vectors as it would create additional threads without assigned work. In the case of matrix vector multiplication we create 1D work groups instead of 2D as we do in the GEMM kernel in order to avoid spawning overhead threads. When multiplying a $1 \times n$ row vector with a $n \times m$ matrix we create $m$ work groups of arbitrary size (default is 64). Each work group is assigned one of the $m$ scalar products.

\subsubsection{Primitive function}

The primitive function for general matrix multiplication accepts two \pkg{Eigen} matrices with elements of type \code{double}. Setting \code{STAN\_OPENCL} flag does not guarantee that all functions that have GPU support will use the GPU. The function will use the existing CPU sequential implementations if the problem is small enough that no speedup is expected. For the primitive \code{GEMM}, the GPU routines are used if $n\times m > 250000$ and $k>100$. These values are set so that the matrix multiplication that meets the criteria is guaranteed to be faster on any given GPU. These values could generally be set lower on mid and high-end GPUs 

\subsubsection{Reverse mode function}

We focused on optimizing the reverse mode for the general matrix multiplication (\code{GEMM}). Reverse modes for multiplication that involve scalars are not optimized using \proglang{OpenCL}. There are three implementations for \code{GEMM} in reverse mode: matrix multiplication of two matrices of \code{stan::math::var} objects, matrix multiplication of a matrix of doubles with a matrix of \code{var} objects and vice versa. The description of the general case is given here, with the other two cases only omitting certain copies.

In the forward pass we extract the values from the matrices and multiply them using the same kernel that is used in the primitive function. If the input matrix is a matrix of doubles, the extraction step can be ignored. In the $chain()$ function, that is used to evaluate the gradient, we have three input matrices of doubles ($adjAB$, $B$, $A$) and we need to calculate the following expressions:

\begin{lstlisting}
adjA = adjAB * B.transpose();
ajdB = A.transpose() * adjAB;
\end{lstlisting}

The \code{transpose} kernel is described in Section \ref{sec:supportRoutines} while the \code{GEMM} kernel explained above is used for the multiplication. The thresholds on when to move the computation to the \proglang{OpenCL} device are the same as for the primitive form of matrix multiplication.

\subsection[Solving triangular systems]{Solving triangular systems}\label{sec:solvingTriSystems}

\subsubsection{Primitive function}

We implemented a general solver for $Ax=b$, where $A$ is triangular and a special case where $b$ is an identity matrix. In the latter, we transfer the input matrix to the GPU, if not already present in the GPUs global memory, and calculate the lower triangular inverse of $A$ on the GPU. If $A$ is upper triangular the input and output of the lower triangular inverse are transposed. The result is then copied back to the global memory of the host. The general solver only adds a multiplication of the inverse $A^{-1}$ with the right-hand side $b$.

\begin{figure}[htp]
\includegraphics[width=8cm]{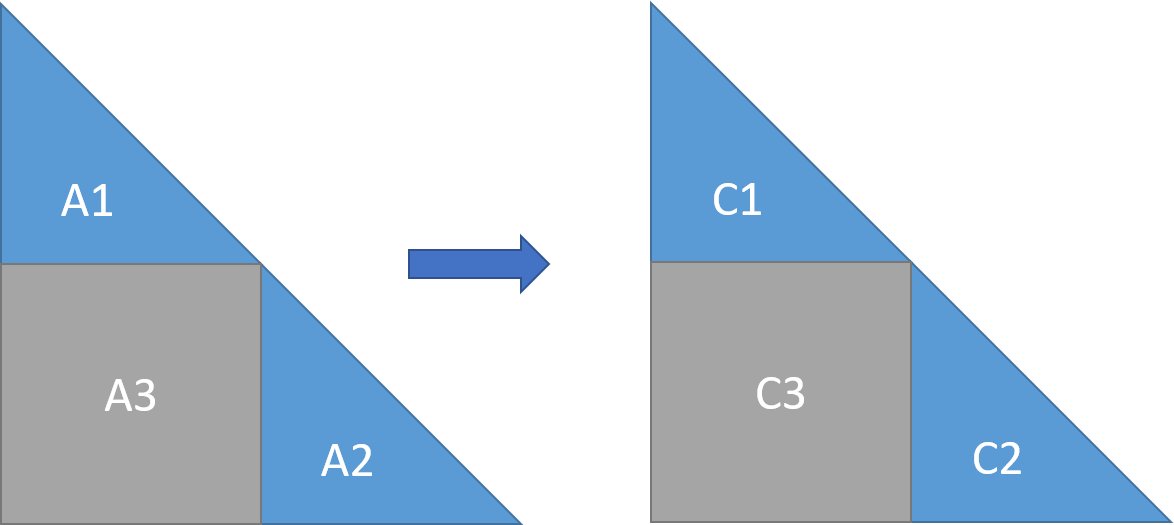}
\centering
\caption{Splitting into submatrices when computing the lower triangular inverse.}
\label{fig:inverseBasic}
\end{figure}

The \proglang{OpenCL} implementation of the lower triangular inverse is based on our previous work \citep{Cesnovar2017}. The forward substitution algorithm that is commonly used in sequential implementations is not suitable even for multi-core parallel implementations as it requires constant communication between threads. The communication-free algorithm for multi-core CPUs proposed in \citep{Mahfoudhi2012} is based on the divide-and-conquer approach. The idea is that we can split the matrix into submatrices as shown in Figure \ref{fig:inverseBasic}. The input matrix is split in three submatrices: $A_1$, $A_2$ and $A_3$. We calculate the inverses of $A_1$ and $A_2$ in parallel to get C1 and C2. The remaining submatrix is then calculated using $C_3 = - C_2A_3C_1$, with matrix multiplication also parallelized as shown in Section \ref{sec:matrixMul}.

\begin{figure}[htp]
\includegraphics[width=8cm]{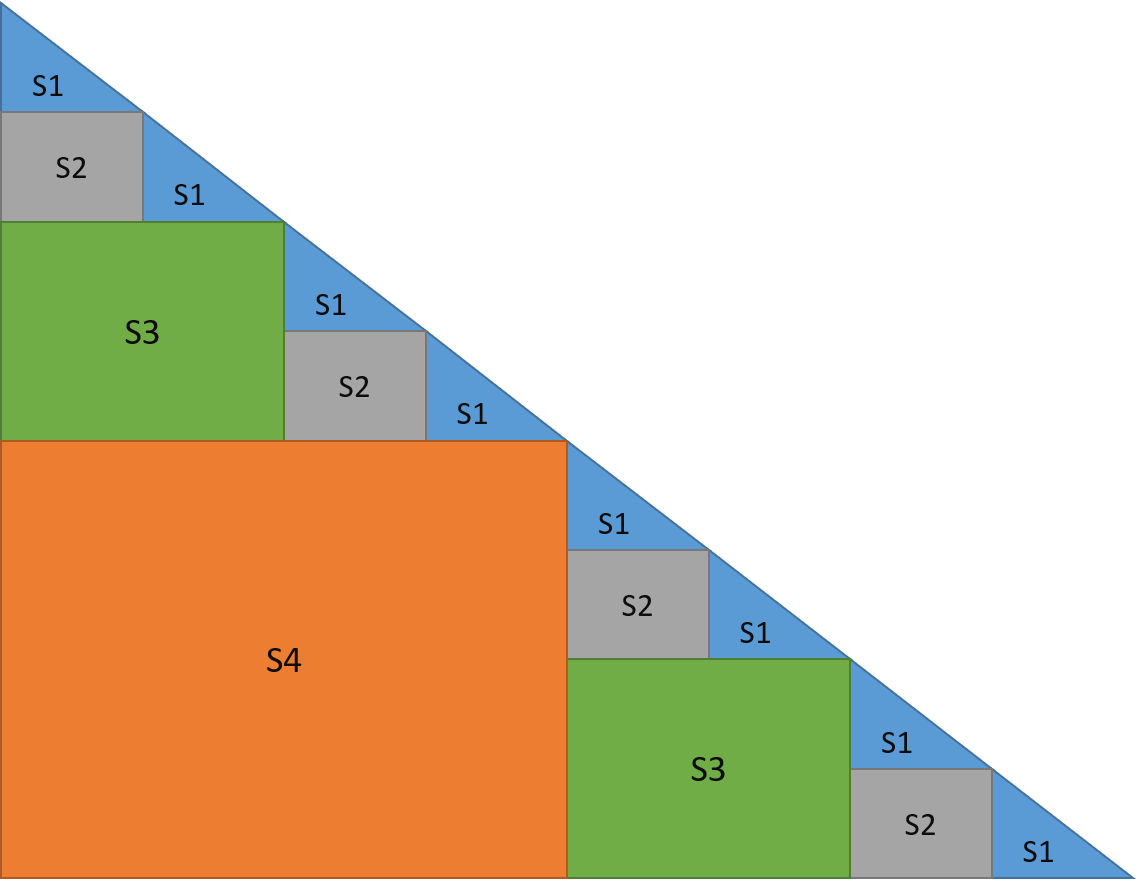}
\centering
\caption{Splitting into submatrices when computing the lower triangular inverse.}
\label{fig:generalInverse}
\end{figure}

Our approach generalizes this for many-core architectures, calculating a batch of smaller inverses along the diagonal in the first step (blocks labeled $S1$ in Figure \ref{fig:generalInverse}). For this step we use kernel \code{batch\_identity} (see \ref{sec:supportRoutines}). It creates a large number of smaller identity matrices that are then used in \code{diag\_inv} to calculate the inverses of the smaller submatrices along the diagonal. For a $n\times n$ matrix, this kernel is executed with $n$ threads split into $b$ work groups, where $b$ is the number of submatrices along the diagonal. Each work group is thus assigned one of the inverses along the diagonal. The rest of the submatrices are calculated by applying equation $C_3 = -C_2 A_3 C_1$. For Figure \ref{fig:generalInverse} this is done in four steps, first calculating submatrices labeled $S_2$, then reapplying the same equation to calculate submatrices $S_3$, $S_4$ and $S_5$. Each step is done in two phases, first calculating $T = C_2A_3$ with the \code{inv\_lower\_tri\_multiply} and then $C_3 = -T C_1$ with the \code{neg\_rect\_lower\_tri\_multiply} kernel. Both kernels are based on the general matrix multiply kernel but modified to handle a batch of multiplications of small submatrices. The GPU support is only used when $n > 500$.

\subsubsection{Reverse mode function}

In order to add GPU support to the reverse mode implementation of triangular system solvers, we used the lower triangular inverse kernel, the \code{GEMM} kernels, and the trivial transpose kernel.

There are again three implementations for solving triangular systems in the form $Ax=b$, one with $A$ and $b$ being matrices of \code{var}, and two cases where either of them is a matrix of doubles. In the function evaluation phase, we use the same principles as in the primitive function with the added steps of extracting a matrix of doubles from the inputs, if needed. In the $chain()$ function that is used to evaluate the gradient, we have three input matrices: $adjC$, $A$ and $C$. To evaluate the gradient we calculate the following: 

\begin{lstlisting}
A * adjB = adjC;
adjA = - adjB * C.transpose();
\end{lstlisting}

Solving the system is done the same way as for the primitive function while the rest is done using a \code{GEMM} kernel, explained in Section \ref{sec:matrixMul}, a transposing kernels and a kernel to multiply a matrix with a scalar, both described in Section \ref{sec:supportRoutines}.

\subsection[Cholesky decomposition]{Cholesky decomposition}

\subsubsection{Primitive function}

We implement fast Cholesky decomposition on the GPU using batched linear algebra, which factors large problems into smaller problems that can be solved in parallel \citep{Abdelfattah2016,Dongarra2016}. The developments in this area have also brought advances for the native mode Cholesky decomposition of a single large matrix. The state-of-the-art implementations presented in \citep{Dong2014, ABDELFATTAH2017,cuBLAS} use \proglang{CUDA} and are limited to NVIDIA GPUs. Our approach is based on the blocked version of the algorithm proposed by \cite{LouterNool1992}. It is based on the idea that if we partition the input matrix as shown in Figure \ref{fig:choleskyBasic}, we can calculate the resulting submatrices as follows:

$$L_{11} = \text{chol}(A_{11})$$
$$L_{21} = A_{21}(L_{11}^T)^{-1}$$
$$L_{22} = A_{22} - L_{21}(L_{21}^T)$$

The function \code{chol} computes the Cholesky decomposition of a smaller $b\times b$ block $A_{11}$ using the classic sequential algorithm. It is implemented with a simple kernel executed with $b$ threads. Parallel execution in the sequential algorithm is limited to parallelizing the inner loop and requires a lot of synchronization so all threads are organized in a single work group. These computations utilize the \code{lower\_triangular\_inverse} from Section \ref{sec:solvingTriSystems} and the matrix multiplication and \code{multiply\_transpose} from Section \ref{sec:matrixMul}. We traverse the input matrix using:

\begin{figure}[tp]
\includegraphics[width=8cm]{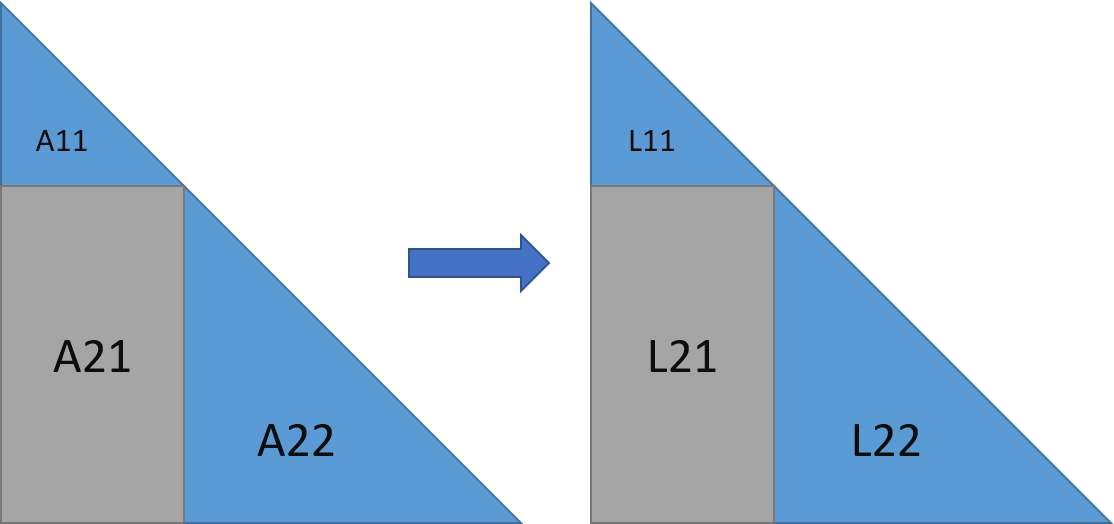}
\centering
\caption{Splitting into submatrices when computing the Cholesky decomposition.}
\label{fig:choleskyBasic}
\end{figure}

\begin{lstlisting}
cholesky_decompose(A) {
  if (A.rows() == 0)
    return A;
 
  if (A.rows() <= min_L11_size) {
	return cholesky_decompose_kernel(A)
  }

  block = A.rows() / partition

  A_11 = A(0:block, 0:block)
  L_11 = cholesky_decompose(A_11);

  A(0:block, 0:block) = L_11

  block_subset = A.rows() - block
  A_21 = A(block:A.rows(), 0:block)
  L_21 = A_21 * transpose(lower_triangular_inverse(L_11))
  A(block:A.rows(), 0:block) = L_21

  A_22 = A(block:A.rows(), block:A.rows())

  L_22 = A_22 - multiply_transpose(L_21)

  L_rem_11 = cholesky_decompose(L_22);
  A(block:N, block:N) = L_rem_11
  return A;
}

\end{lstlisting}

Note that \code{partition} and \code{min_L11_size} are implementation parameters.

\subsubsection{Reverse mode function}

We re-implemented the \pkg{Stan Math} blocked Cholesky decomposition gradient in \proglang{OpenCL}. The blocked version of the gradient is based on the work of \cite{Murray2016}. The input to the gradient is the values and adjoints of the reverse mode input matrix. Both matrices are lower triangular so their values are copied to the GPU in a packed form and then unpacked with the \code{unpack} kernel (see Section \ref{sec:supportRoutines}). The resulting values of the adjoints are packed with the \code{pack} kernel and transferred back to the host's global memory. The gradient calculation pseudo-code is given below:

\begin{lstlisting}
for (k = N; k > 0; k -= block_size_) {
      j = max(0, k - block_size_);
      R = L(j:k, 0:j)
      D = L(j:k, j:k)
      B = L(k:N, 0:j)
      C = L(k:N, j:k)
      R_adj = L_adj(j:k, 0:j)
      D_adj = L_adj(j:k, j:k)
      B_adj = L_adj(k:N, 0:j)
      C_adj = L_adj(k:N, j:k)

      C_adj = C_adj * lower_triangular_inverse(D)
      B_adj = B_adj - C_adj * R;
      D_adj = D_adj - transpose(C_adj) * C;

      D_adj = transpose(D) * D_adj
      copy_lower_tri_to_upper_tri(D_adj)
      D = transpose(lower_triangular_inverse(D));
      D_adj = D * transpose(D * D_adj)
      copy_lower_tri_to_upper_tri(D_adj)

      R_adj = R_adj - transpose(C_adj) * B - D_adj * R;
      D_adj = D_adj.diagonal() * 0.5
      set_zeros_in_upper_tri(D_adj)
}
\end{lstlisting}

Note that \code{block_size} is set heuristically with default values that can be tuned for target GPUs.

\subsubsection{Speedup measurements}

\begin{figure}
\includegraphics[width=\linewidth]{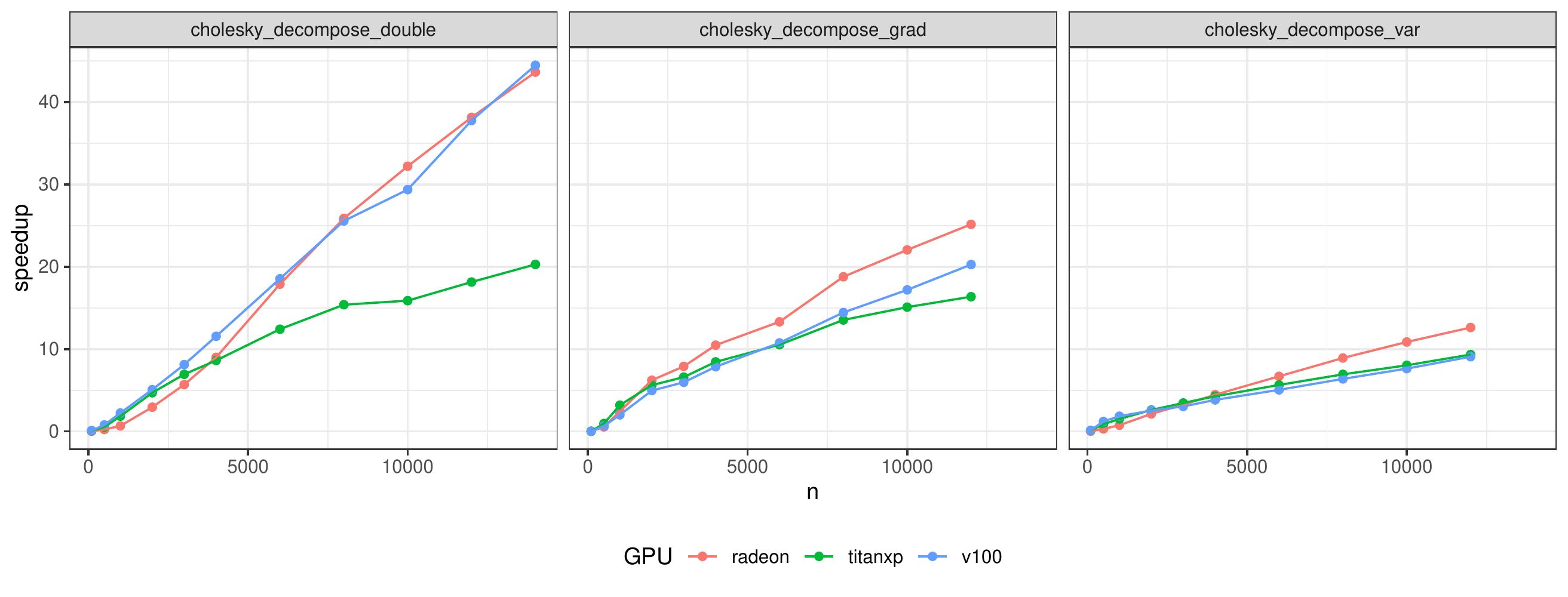}\caption{Cholesky decomposition speedups for different GPUs and varying input matrix size $n \times n$. We measured the performance of the Cholesky decomposition on matrices of doubles, the gradient of the Cholesky decomposition, and matrices of \code{var}. The latter are required when the matrix is a parameter, for example, in Gaussian Process regression.}\label{fig:cholesky}
\end{figure}

We measured computation times of the Cholesky decomposition on an Intel Core i7-5600U CPU at 3.60GHz and three different GPUs: NVIDIA Titan XP (titanxp), NVIDIA Tesla V100 (v100), and AMD Sapphire Radeon VII (radeon). We ran each experiment for different input matrix sizes in increments of 1000 up to the size that caused a \code{std::bad_alloc} or out of memory on the GPU. We generated Toeplitz matrices with $A_{i,j; i \neq j} = n - |i - j|$ and $A_{i,i} = n^2$ to ensure positive definiteness. The measurement used at each input size was the median of 9 repetitions.  Results are shown in Figure \ref{fig:cholesky}.

\subsection[Generalized linear models]{Generalized linear models}

Generalized linear models (GLMs) are a popular family of statistical models. A GLM is characterized by $\text{E}[Y] = g^{-1}(X\beta$), where $Y$ is the target variable, $X\beta$ is a linear function of input variables and coefficients, and $g$ is the link function that connects the linear term with the mean of the dependent variable.

Computing the likelihood for a GLM consists of computing the linear combination, transforming it with $g^{-1}$ and finally computing the likelihood according to our choice of distribution for $Y$. We can improve performance by computing all steps simultaneously (including analytically derived gradient), instead of performing each step individually and relying on automatic differentiation for the gradient. Six such likelihood primitives are implemented in the \pkg{Stan Math} library: normal-identity (normal distribution with identity link, linear regression), Bernoulli-logit (logistic regression), Poisson-log, negative binomial-log, categorical-logit (multinomial logistic regression), and ordinal-logit (ordinal logistic regression). \proglang{Stan} users can access them by calling the built-in log-pdf/pmf. These likelihoods can also be used as components in more complex models, such as multi-level regression.

Note that the GLM likelihoods are heavily templated. With the exception of integers, every argument can have type \code{var} or \code{double}. Many arguments can either be scalars or vectors. For example, standard deviation in the normal-identity GLM can be the usual scalar parameter or, if we want to implement heteroskedasticity, a vector.

The GPU implementation of the GLM primitives is based on their CPU implementation that is already in \proglang{Stan} and not part of this work. First, the data are transferred to the GPU. The argument types also determine which derivatives we need to compute. This information is transferred to the GPU kernel by kernel arguments and it allows kernels to skip unnecessary computation. 

Each GLM is implemented in a single kernel. The kernel requires execution of one thread per input variable. The number of threads is rounded up and they are organized into work groups of size 64. Each thread calculates one scalar product of the matrix-vector product and additional GLM-specific values and derivatives. Computation of the (log) likelihood ends with the summation over intermediate values, computed by each thread. Threads in a work group execute parallel reduction. First thread of each work group writes partial sum to the global memory. Partial sums are then transferred to the main memory and summation is completed by the CPU. 

Derivatives with respect to coefficients and input variables require calculation of another matrix-vector product. Since another product cannot be efficiently computed in the same kernel, a \code{GEMV} kernel is run if these derivatives are needed.

\subsection[Supporting GPU routines]{Supporting GPU routines}\label{sec:supportRoutines}

Here we describe additional \proglang{OpenCL} kernels that are available in \pkg{Stan Math}. These routines are not bottlenecks of statistical computation, so we opted for simplicity. All kernels use the default work group size determined by \proglang{OpenCL}.

These \proglang{OpenCL} kernels often do not outperform their CPU or \pkg{Eigen} equivalents, especially when we consider data transfers to or from the \proglang{OpenCL} GPU device. They should only be used as parts of an algorithm or when there is no data copy overhead. Note that all \proglang{OpenCL} kernels in \pkg{Stan Math} can be used as building blocks for new GPU routines.

All the kernels below assume input matrices of size m$\times$n.

\subsubsection[Add and subtract]{Add and subtract}

These kernels add or subtract two matrices of the same size and store the result to a third matrix buffer. They execute on a grid of $m\times n$ threads, where thread $(i,j)$ adds or subtracts elements $A_{i,j}$ and $B_{i,j}$ and stores them to $C_{i,j}$.

The kernel code for adding two matrices:

\begin{lstlisting}
__kernel void add(__global double *C, __global double *A,
                      __global double *B, unsigned int rows,
                      unsigned int cols) {
        int i = get_global_id(0);
        int j = get_global_id(1);
        if (i < rows && j < cols) {
          C(i, j) = A(i, j) + B(i, j);
      }
    }
\end{lstlisting}

\subsubsection[Multiplication with scalar]{Multiplication with scalar}

Kernels \code{scalar\_mul} and \code{scalar\_mul\_diagonal} can be used to multiply a matrix with a scalar. The former multiplies the entire input matrix with the given scalar while the latter multiplies only the diagonal elements. Similarly to the add and subtract kernels, the kernel \code{scalar\_mul} is executed with $m\times n$ threads with each thread multiplying one matrix element. The kernel \code{scalar\_mul\_diagonal} is executed with $\min(m,n)$ threads with each thread multiplying one diagonal element.

\subsubsection[Matrix transpose]{Matrix transpose}

Two kernels can be used for transposing a matrix and both are executed with $m\times n$ threads. In kernel \code{transpose} each thread copies an element from $A_{i,j}$ to $B_{j,i}$ and \code{triangular\_transpose} copies the lower triangular of a matrix to the upper triangular or vice-versa. In the former case, the threads with indices under and on the diagonal perform $A_{j,i} = A_{i,j}$ and threads above the diagonal do nothing. In the latter case the roles are reversed.

\subsubsection[Matrix initialization]{Matrix initialization}

Three kernels can be used to initialize a matrix: \code{zeros}, \code{identity} and \code{batch\_identity}. For kernel \code{zeros} we specify the output matrix, its size, and if we want to set zeros only on the lower or upper triangular of the matrix. In both cases, we spawn $m\times n$ threads, with each thread assigned a single matrix element. Similarly, for kernel \code{identity} each thread is assigned a single matrix element. The \code{batch\_identity} kernel is used to make a batch of smaller identity matrices in a single continuous buffer. This kernel is executed with $k\times n \times n$ threads, where $k$ is the number of $n\times n$ identity matrices to create. Each thread is assigned a single element in one of the batch matrices, with the thread ID $(0, ..., k-1)$ on the first dimension determining the matrix and the IDs on the remaining two dimensions determining the element of the matrix.

\subsubsection[Submatrix copy]{Submatrix copy}

\proglang{OpenCL} provides functions that enable copying of entire matrix buffers. In order to add the functionality of copying arbitrary submatrices we implemented the \code{sub\_block} kernel. This kernel copies a rectangular or triangular submatrix of size $k\times l$ from a source matrix to a given submatrix of the destination matrix of size $m\times n$. Each of the $k\times l$ threads that execute this kernel is assigned one element of the submatrix. The thread then determines whether to copy the element based on the triangular view argument and its indices on the two dimensions. Similarly to the \code{zeros} kernel, \code{sub\_block} has a triangular view argument that determines whether to copy the entire input matrix or only the lower/upper triangular parts.

\subsubsection[Input checking]{Input checking}

We implemented kernels that allow the user to check whether the matrices stored on the \proglang{OpenCL} devices are symmetric, contain \code{NaN} values or contain zeros on the diagonal. The inputs are a pointer to the matrix buffer, its size and a pointer to a flag variable. If the conditions of the check are met, the matrix sets the flag. No threads reset the flag so there are no race conditions. \code{check\_diagonal\_zeros} is executed with $\min(m, n)$ threads, each thread checking one diagonal element for zeros. \code{check\_nan} is executed with $m\times n$ threads, each thread checking one matrix element for \code{NaN}. For the kernel \code{check\_symmetric} the flag should be set before executing the kernel with the threads resetting it upon discovery of non-symmetry. The kernel is again executed using $m\times n$ threads, each thread checking whether $A_{i,j} - A_{j,i}$ is within tolerance of zero.

\subsection[Kernel fusion]{Kernel fusion}\label{sec:kf}

If only the basic computation primitives are implemented for the GPU, we are faced with a dilemma when developing new algorithms that are a composition of these primitives (for example, GLM likelihoods). We can call multiple kernels or manually develop a new kernel that combines multiple primitive methods into a single kernel. The latter speeds up computation in two ways. First, we reduce kernel launch overhead. And second, we reduce the amount of memory transfers. Combining kernels is also referred to as \textit{kernel fusion} \citep{filipovic2015} and it can be done manually or automated.

The problem with manually developing new kernels is that it requires specific expertise and is a time consuming process. To mitigate this, we also developed an OpenCL GPU kernel fusion library for \pkg{Stan Math}. The library automatically combines kernels, optimizes computation, and is simple to use. This speeds up the development of new GPU kernels while keeping the performance of automatically combined kernels comparable to hand crafted kernels. The details of the kernel fusion library for \pkg{Stan Math} can be found in \cite{Ciglaric2019}.

We maintain a list of functions that are implemented and available for use with the kernel generator at \url{https://github.com/bstatcomp/math/issues/15}. It includes the previously mentioned add and subtract, multiplication with scalar, and submatrix copy.  

\section[Illustrative examples]{Illustrative examples} \label{sec:illustrations}

In this section we show two illustrative examples that demonstrate the use of the new GPU routines and what speedups we can expect.

\begin{itemize}
\item The measurements are end-to-end, measuring the speedups that a user would experience during typical use. That is, we measure the time it takes from calling the model to sample to when the samples are output by the model. This includes reading the data and outputting the samples, but not model compilation time. Model compilation only has to be done once and we find it reasonable to assume that compilation time is negligible over the life span of the model. For logistic regression we ran each experiment for 500 warmup and 500 sampling iterations and for GP regression 30 warmup and 10 sampling iterations.
\item We made three measurements for each hardware-model-input size configuration. We calculate speedups by dividing the average of the three measurements for the device with the average of the three measurements for the reference device (cpu).
\item We ran the experiments on an Intel Core i7-5600U CPU (cpu) at 3.60GHz and two different GPUs: NVIDIA Titan XP (titanxp), and AMD Sapphire Radeon VII (radeon).
\end{itemize}

\subsection[Logistic regression]{Logistic regression}\label{sec:logistic}

We generated toy datasets as follows:

\begin{itemize}
\item Each element of the $n\times k$ matrix $X$ is an independent draw from the standard normal. Dimensions $n$ and $k$ will vary.
\item The latent linear term $\mu_i$ depends only on the first two input variables $\mu_i = 3 X_{i,1} - 2 X_{i,2} + 1$. The remaining $k - 2$ input variables are noise.

\item Target variable $y_i$ is set to 1 with probability $\frac{1}{1 + e^{-\mu_i}}$.
\end{itemize}

The model is standard logistic regression with Bernoulli likelihood and logit link function:

$$y_i|\alpha, \beta \sim_\text{iid} \text{Bernoulli}\left(\frac{1}{1 + e^{-( X_i\beta + \alpha)}}\right).$$

\begin{figure}[htp]
\centering
\subfloat[][times ($k$ = 10)]{
\includegraphics[width=0.47\textwidth]{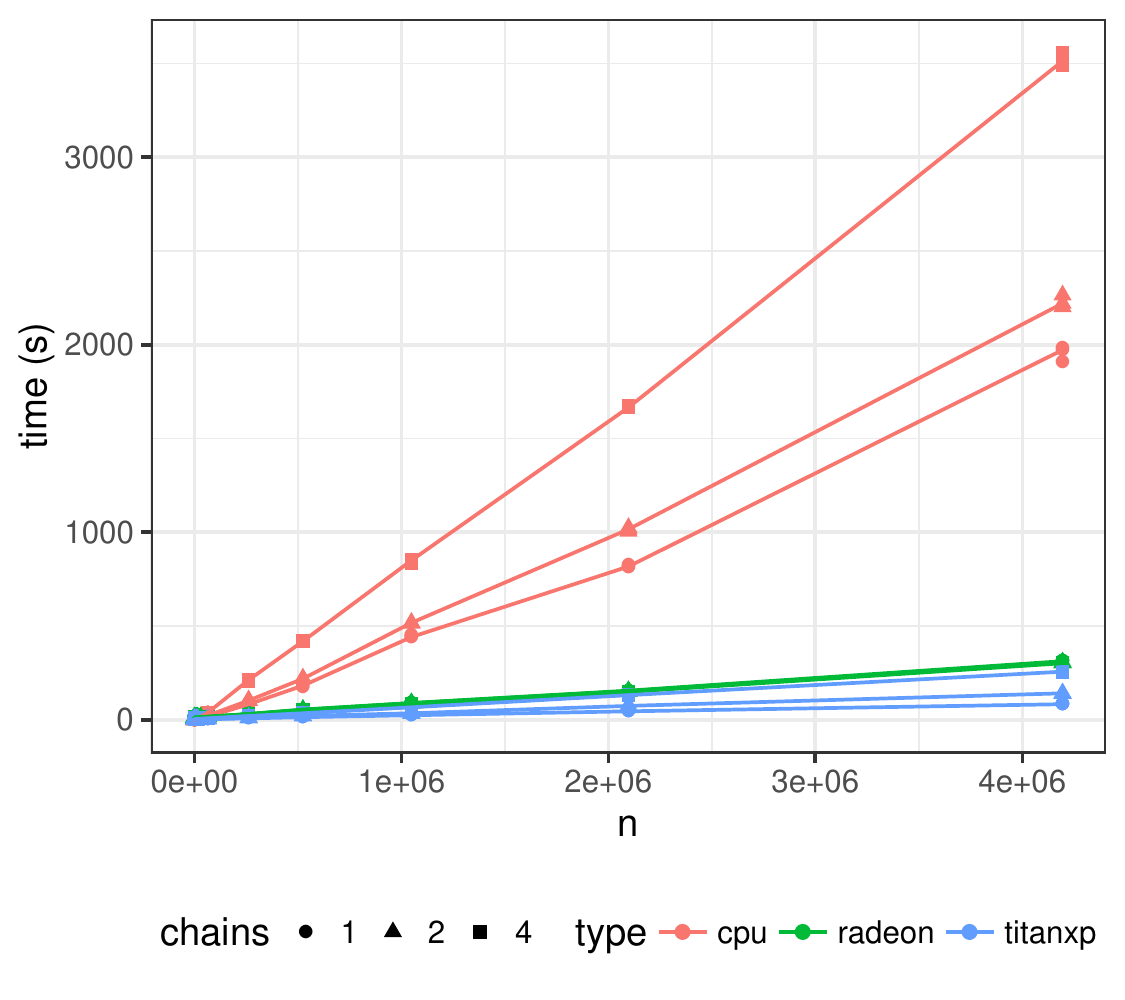}
\label{fig:subfig1}}
\subfloat[][speedups ($k$ = 10, 4 chains only)]{
\includegraphics[width=0.47\textwidth]{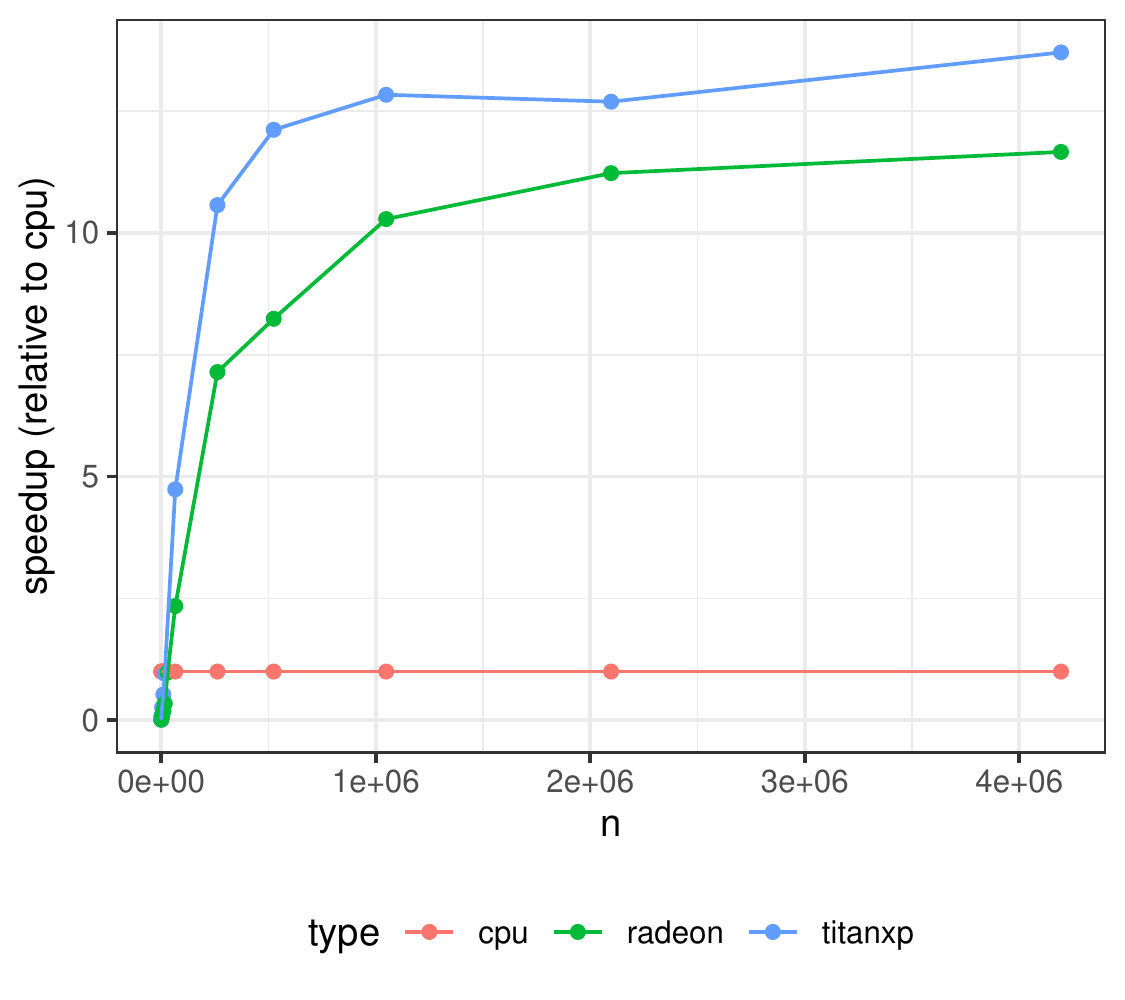}
\label{fig:subfig2}}
\qquad
\subfloat[][times ($n$ = 10000)]{
\includegraphics[width=0.47\textwidth]{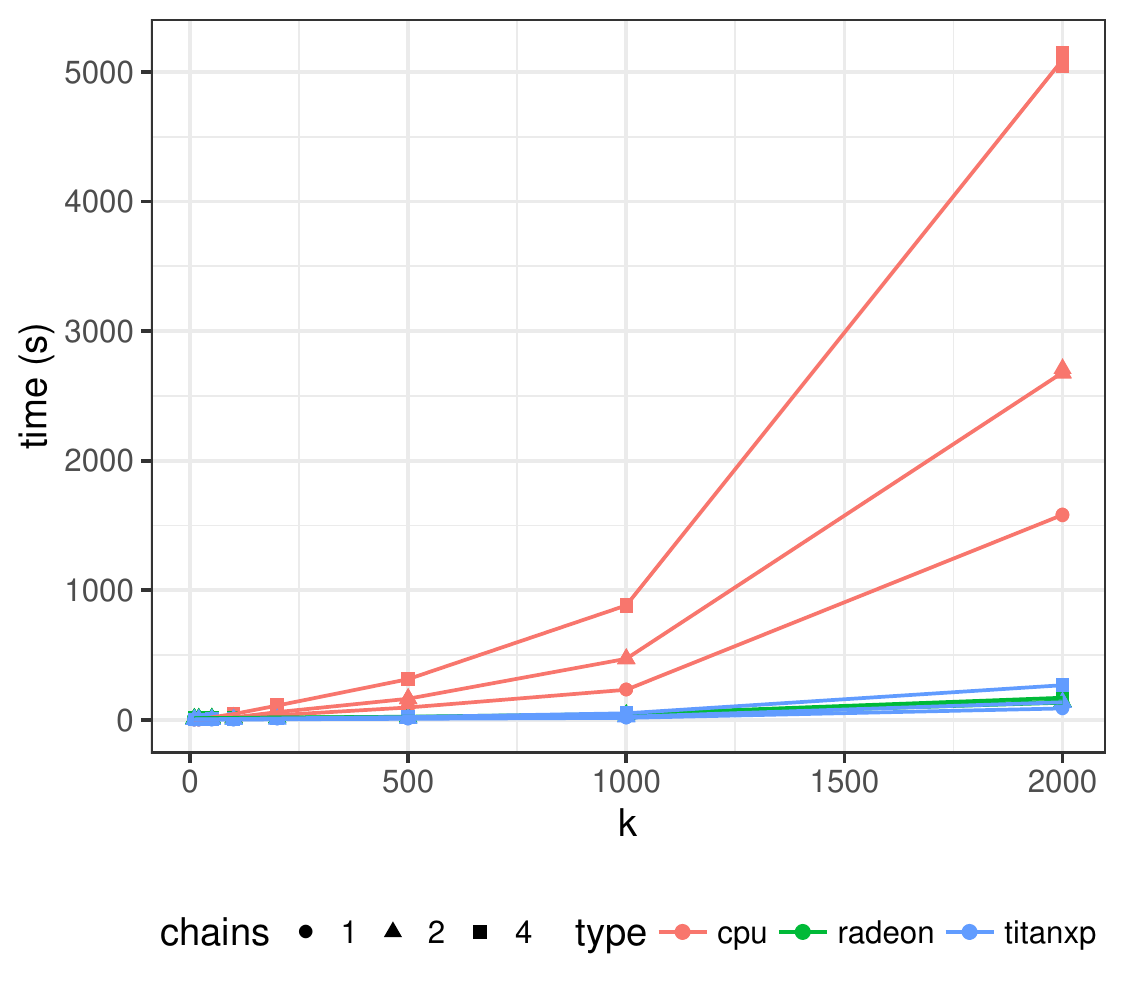}
\label{fig:subfig3}}
\subfloat[][speedups ($n$ = 10000, 4 chains only)]{
\includegraphics[width=0.47\textwidth]{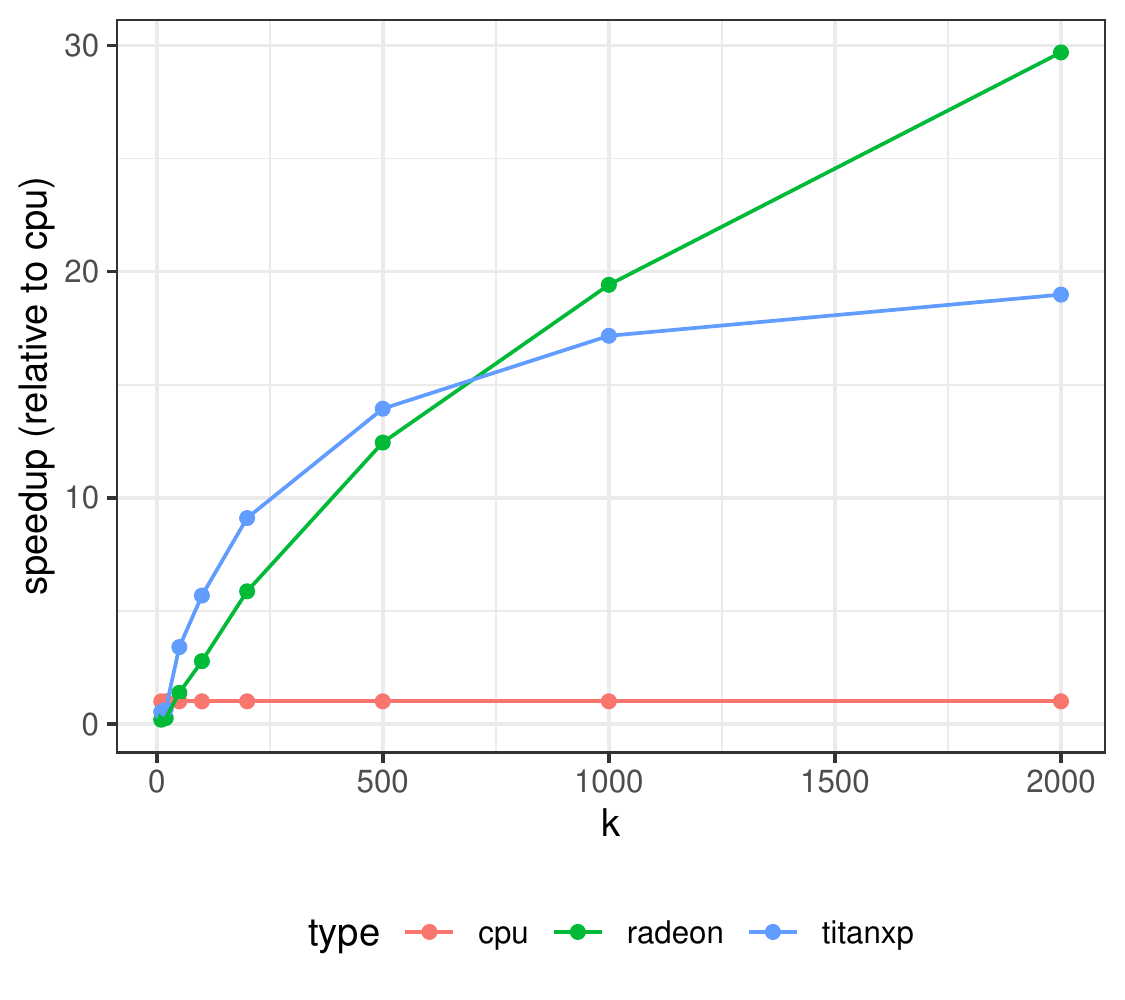}
\label{fig:subfig4}}
\caption{Times and speedups for logistic regression. Note the that the number of observations $n$ and the number of input variables $k$ were limited by CPU computation time. Some plots indicate that GPU performance has not yet peaked -- better speedups can be expected for larger $n$ and/or $k$. Also note that running 4 chains in parallel takes less time than running 1 chain 4 times. This is true for CPU-only and GPU. With multiple chains we can better utilize the GPU.}
\label{fig:reslogistic}
\end{figure}

We used the default flat priors provided by \proglang{Stan}\footnote{The default flat prior is improper and will lead to an improper posterior if the data are separable. However, in our toy data set the data are not separable and the model is used only to illustrate the speedups.}:

\begin{lstlisting}
data {
  int<lower=1> k;
  int<lower=0> n;
  matrix[n, k] X;
  int y[n];
}

parameters {
  vector[k] beta;
  real alpha;
}

model {
  target += bernoulli_logit_glm_lpmf(y | X, alpha, beta);
}
\end{lstlisting}

The results are summarized in Figure~\ref{fig:reslogistic}.

\subsection[Gaussian Process regression]{Gaussian Process regression}

We generated toy datasets as follows:

\begin{itemize}
\item Each of the $n$ elements of $x$ is an independent draw from $\text{Unif}(-10,+10)$. Dimension $n$ will vary.
\item Target variable $y_i$ is drawn from $\text{N}( f(x), \frac{1}{10})$, where $f(x) = \beta(x + x^2 - x^3 + 100 \sin 2x - \alpha)$. Parameters $\beta$ and $\alpha$ were set so that $\text{E}[f] = 0$ and $\text{Var}[f] = 1$. 
\end{itemize}

The model is a 1D Gaussian Process regression with hyper-priors as implemented by \cite{Betancourt2017}. For the sake of brevity, we do not include the model code in the text.

\begin{figure}[htp]
\centering
\subfloat[][times (variable $n$)]{
\includegraphics[width=0.47\textwidth]{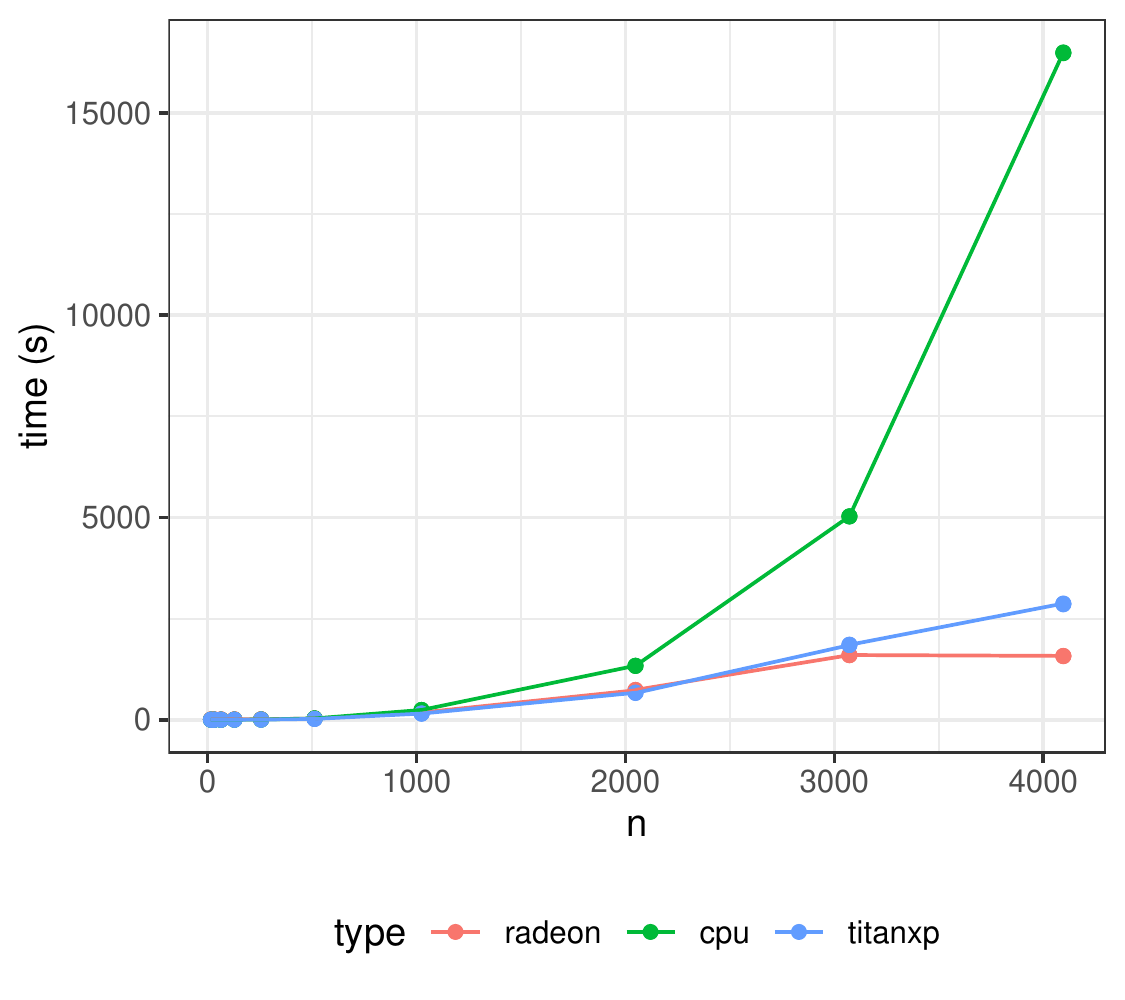}
\label{fig:subfig1}}
\subfloat[][speedups (variable $n$)]{
\includegraphics[width=0.47\textwidth]{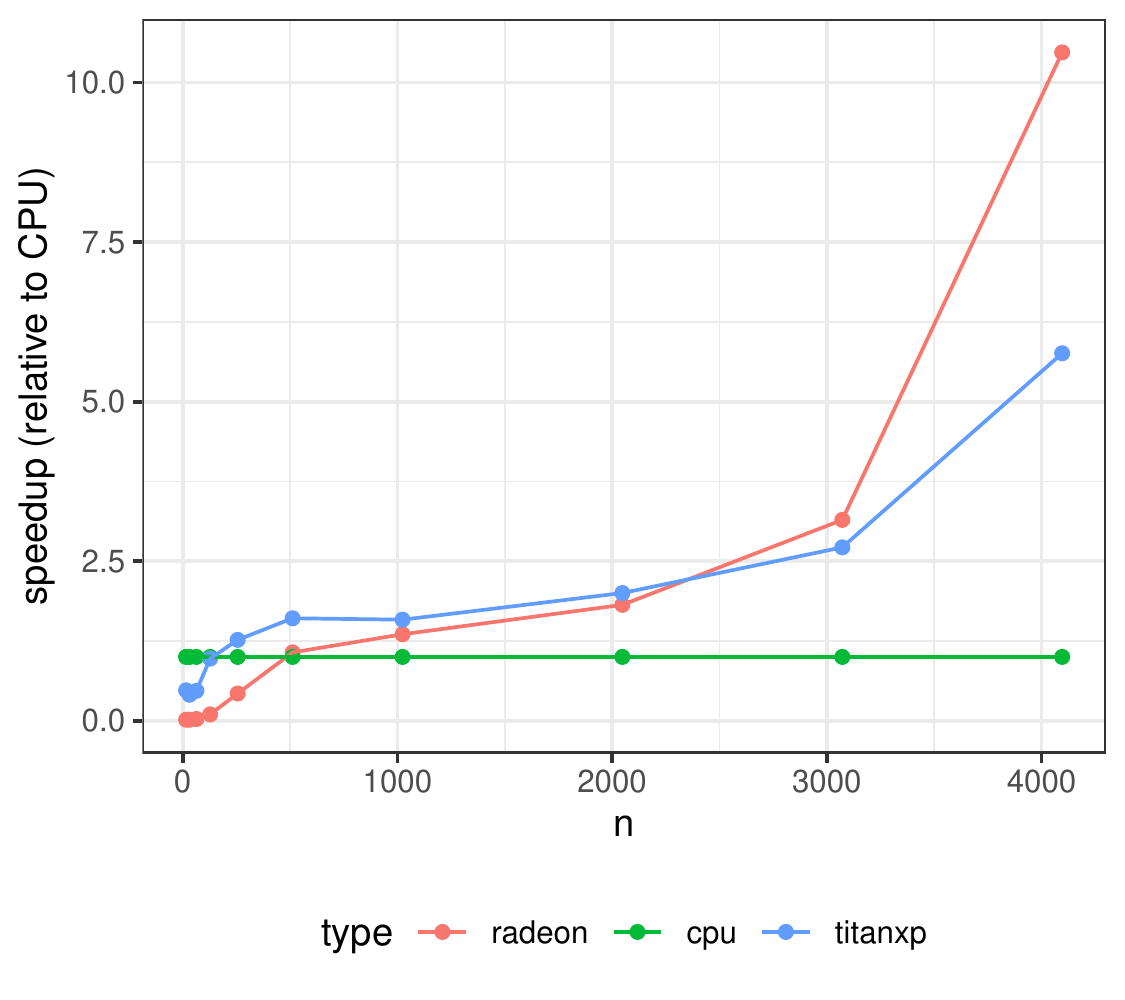}
\label{fig:subfig2}}
\caption{Times and speedups for Gaussian Process regression. Again, the number of observations $n$ was limited by CPU computation time - better speedups could be achieved for higher larger $n$. The plots indicate that GPU performance has not yet peaked for these GPUs -- better speedups can be expected for larger $n$.}
\label{fig:resgp}
\end{figure}

The results are summarized in Figure~\ref{fig:resgp}.

\section{Discussion and conclusion} \label{sec:conclusion}

GPU support provides \proglang{Stan} users with practically meaningful speedups in moderate to large problems, even on mid-range retail gaming GPUs. Furthermore, the speedup estimates from our illustrative examples are very conservative for two reasons. First, additional speedup can be achieved by using compute-only GPUs. And second, a substantial part of the time in the logistic regression example is due to slow I/O in \proglang{CmdStan} -- the data (samples) are passed (from) the model as text files and their size grow linearly with the number of input variables and observations as does the computation of these likelihoods.

The benefits come with zero user effort beyond setting up a GPU and without the need to change existing \proglang{Stan} model code. This is an ongoing project - we will be adding other matrix algebra and statistical likelihood primitives, such as other matrix decompositions, covariance functions, and likelihoods. Expert users can already take advantage of the API and add their own GPU kernels.

The API for \proglang{Stan}'s \proglang{OpenCL} backend is still evolving. The next major step is to add support for multiple heterogeneous devices and an efficient load balancing system. Tuning such computation is a challenge. It consists of two parts: (a) how to set optimal parameters for the device(s) and (b) when to move computation to the device(s) (when does the speedup justify the overhead). The iterative Markov Chain Monte Carlo (or optimization) setting lends itself to the possibility of efficient online tuning, because computation and input dimensions are constant over all iterations.

\section*{Computational details}

All the functionality described in this paper is part of Stan as of releases \pkg{CmdStan}~2.22 and \pkg{Stan Math}~3.1. 

Instructions on how to activate GPU support for CmdStan or CmdStanR can be found here: \url{https://github.com/bstatcomp/stan_gpu_install_docs}

\section*{Acknowledgments}

We would like to thank Bob Carpenter for for his comments on an earlier draft of our work. This research was supported by the Slovenian Research Agency (ARRS, project grant L1-7542 and research core funding P5-0410). We gratefully acknowledge the support of NVIDIA Corporation with the donation of the Titan XP GPU used for this research. We gratefully acknowledge the support of Amazon with an Amazon Research Award. Part of Steve Bronder's contributions were made while he was working at Capital One.


\bibliography{main}

\newpage

%
%
%

\end{document}